\documentclass{aastex}
\usepackage{emulateapj5,psfig}
\usepackage{natbib,psfig}
\usepackage{graphicx}

\newcommand{\gtsim}{\mbox
{{\raisebox{-0.4ex}{$\stackrel{>}{{\scriptstyle\sim}}$}}}}
\newcommand{\ltsim}{\mbox
{{\raisebox{-0.4ex}{$\stackrel{<}{{\scriptstyle\sim}}$}}}}

\slugcomment{Accepted by ApJ Letters}

\shorttitle{A relativistic jet in a radio-quiet quasar}
\shortauthors{Blundell, Beasley \& Bicknell}

\begin{document}

\title{A relativistic jet in the radio-quiet quasar PG\,1407+263}

\author{Katherine M.\ Blundell\altaffilmark{1}, Anthony J.\
Beasley\altaffilmark{2} and Geoffrey V. Bicknell\altaffilmark{3}}

\altaffiltext{1}{University of Oxford, Astrophysics, Keble Road, Oxford,
OX1 3RH, UK}
\altaffiltext{2}{Owens Valley Radio Observatory, California Institute
   of Technology, PO Box 968, Big  Pine, CA 93513}
\altaffiltext{3}{Research School of Astronomy \& Astrophysics,
Australian National University, Canberra, Australia }

\begin{abstract}
We present the results of a multi-epoch radio monitoring campaign
measuring the milliarcsecond structure of the jet in the radio-quiet
quasar PG\,1407+263.  This is the highest-sensitivity,
highest-resolution multi-year study of a distant active galaxy.  The 
observations are naturally explained in terms of a beamed relativistic
jet, some of whose fluctuations in flux density can be ascribed to
interaction with the narrow-line region of the quasar. The optical
properties of PG\,1407+263, in particular the low equivalent widths of
the emission lines, may be related to the fact that we are viewing
this quasar almost pole-on, giving us a direct view into its
broad-line region.
\end{abstract}

\keywords{galaxies: jets --- quasars: individual (PG1407+263) ---  radio
continuum: galaxies}  

\section{Introduction}
\label{sec:intro}

In recent years, the weak radio core emission associated with the
class of so-called `radio-quiet' quasars \citep[for
example,][]{Kel89,Mil90,Mil93} has been shown by a succession of Very
Long Baseline Interferometry (VLBI) imaging experiments usually to be
compact on milliarcsecond (mas) scales \citep{Blu96,Fal96,Blu98}.  The
typical physical size of \ltsim\ a few cubic parsecs for these radio
luminosities and corresponding brightness temperatures ($> 10^6$\,K)
strongly argues against star-formation as the origin of this radio
emission, despite the striking continuity of the correlation of
far-infra-red and radio emission from star-forming galaxies to
radio-quiet quasars \citep{Sop91a,Sop91b}.  Instead, a mechanism
resembling that which powers the core emission in the more
radio-luminous `radio-loud' quasars is suggested.

One quasar in particular, PG\,1407+263, identified in a mas imaging
survey of radio-quiet quasars \citep{Blu98} clearly has multiple
components and seemed an important target to monitor subsequently with
VLBI techniques.  It was first discovered as a quasar in the
Palomar-Green survey \citep{Sch83} with $z = 0.94 \pm 0.02$
\citep{McD95}.  With a Hubble constant of $70 \> \rm km\,s^{-1}
Mpc^{-1}$, and $(\Omega_{\mathrm M},\Omega_{\Lambda}) = (0.3,0.7)$,
the luminosity distance $\approx 6.2 \> \rm Gpc$ and 1\,mas $\approx$
7.9\,pc.

\section{Measurements}
\label{sec:meas}
We observed PG\,1407+263 at 8.4\,GHz using the Very Long Baseline
Array (VLBA) with the Very Large Array (VLA) in phased array mode on
1996 June 15 (VLBA only), 1997 April 27, 1998 January 31, 1999 March
20 and 1999 July 20.  We usually recorded a 64-MHz bandwidth (a 32-MHz
bandwidth was used for epochs 4 \& 5) with 2--bit sampling and dual
circular polarization.  The observations were phase-referenced
\citep{Bea95}, i.e.\ regular observations of a bright extragalactic
source close on the sky to the target were made to derive
residual antenna-based and atmospheric phase corrections, which were
interpolated to calibrate the target data and provide accurate
astrometric referencing.  The calibrator source used was OQ\,208,
separated from PG\,1407+263 by 2.2$^{\circ}$.  We typically used a
switching cycle of two minutes on the calibrator and three minutes on
the target.  The total on-source time for OQ\,208 was 0.5\,hrs (epoch
1) and 2.8\,hrs (epochs 2, 3, 4 and 5).  Throughout the three-year
monitoring, a pair of \ltsim\ mJy components were persistently detected
(Figure\,1), albeit with changes in structure and flux density
(Figure\,2).

\section{Light curves and structural changes}
\label{sec:lightcurves}
One striking change seen between successive epochs is that the
structure of each component appears to vary: this is particularly
marked in the final three epochs (Figure\,1).  Real changes in shape
represent an intriguing aspect of the parsec-scale evolution of the
source: since 1~mas corresponds to 7.9~pc, real expansion between
epochs of one tenth of a synthesized beam would correspond to apparent
motion of a few times the speed of light.  The components appear both
resolved and unresolved at different epochs.  Component B appears to
be significantly resolved (with a high total flux density) in earlier
epochs, while both components appear essentially unresolved (with low
flux densities) in the final epoch.  Unmodeled residual spatial
gradients in tropospheric (and ionospheric) delays lead to
interferometer phase errors which are cancelled towards the bright
calibrator source OQ\,208, but remain in the target data.  Thus, apart
from the above general assertions above about component B, it is
difficult to draw solid conclusions from the structural variations:
simple centroid fitting to the components, while technically possible,
can be ambiguous in interpretation.

The rms variation of the position of component A is $\approx
300$\,$\mu$-arcsec.  Typical estimates of the effects of tropospheric
delay errors on phase-referenced VLBI data with a calibrator 2.2\,deg
away are 100--200\,$\mu$-arcsec. In addition, the typical
signal-to-noise (S/N) of the detections is 5--10 to 1; this would
introduce a positional variation roughly equal to the synthesized beam
divided by the S/N -- i.e.\ 100 to 250\,$\mu$-arcsec.  Therefore, the
position wandering largely results from a combination of unmodeled
tropospheric delay errors and S/N considerations.  Given the weak
(\ltsim\ mJy) flux density levels in these images, it is not possible
to correct for these residual errors.  During the course of these
observations the relative position shift of components A and B
appeared at times to be systematic, possibly indicating superluminal
motion \citep{Blu98b}. Analysis of the existing data set does not
allow a definitive determination of superluminal motion in this
radio-quiet quasar, although we in any case infer it to have a highly
relativistic jet speed (Doppler factor $\gtsim 10$) in the analysis
below.  The rapid variations in component B's flux density are
unlikely to result from imaging coherence loss, which would affect
both components; A's relatively constant peak flux density implies
that B's flux density variations are intrinsic.  The dramatic changes
in B's flux density occur on timescales of 100--300 days, reminiscent
of those seen in the cores of radio-loud quasars \citep{Hou99}.

\psfig{figure=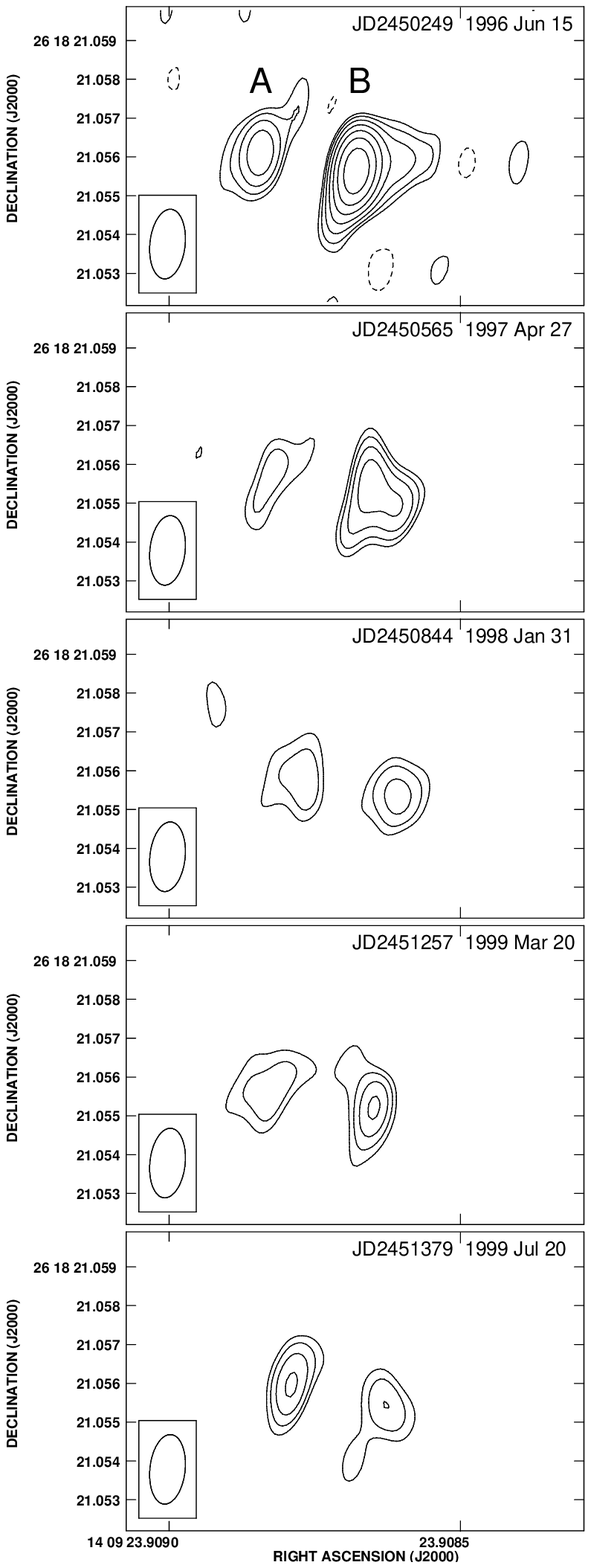,width=7cm,angle=0}
\figcaption{\label{fig:contours} VLBA images of PG\,1407+263 restored
with the same synthesized beam, 1.8 $\times$ 0.9\,mas with position
angle 170\,deg East of North.  The contours in each case are $\pm 0.3$,
0.42, 0.6, 0.84, 1.2, 1.68, 2.4\,mJy/beam.  }

\section{Recent radio-loudness of PG\,1407+263}
\label{sec:radioloud}

In order to establish whether PG\,1407+263 is truly a radio-quiet
quasar, we made deep VLA images to search for evidence of extended
radio structure or hotspots.  These images revealed only independent
sources including an apparent double 53\,arcsec east of the quasar.
\citet{Kel89} commented that this double structure might be related to
the output of the quasar, since it lacks any optical identification on
the Palomar Observatory Sky Survey.  We made a more sensitive test by
observing the field in the near infra-red $K$-band on the UK Infra-Red
Telescope (UKIRT).  Figure\,3 indicates each part of the `double'
radio structure is identified with a galaxy and is thus unrelated to
the quasar.  Therefore, we do not believe that PG\,1407+263 has been
radio-loud for the last $10^{6 - 7}$\,years, where this timescale is
taken from the typical radiative lifetimes for synchrotron-emitting
particles in the lobes of radio-loud quasars \citep{Blu00}.

\psfig{figure=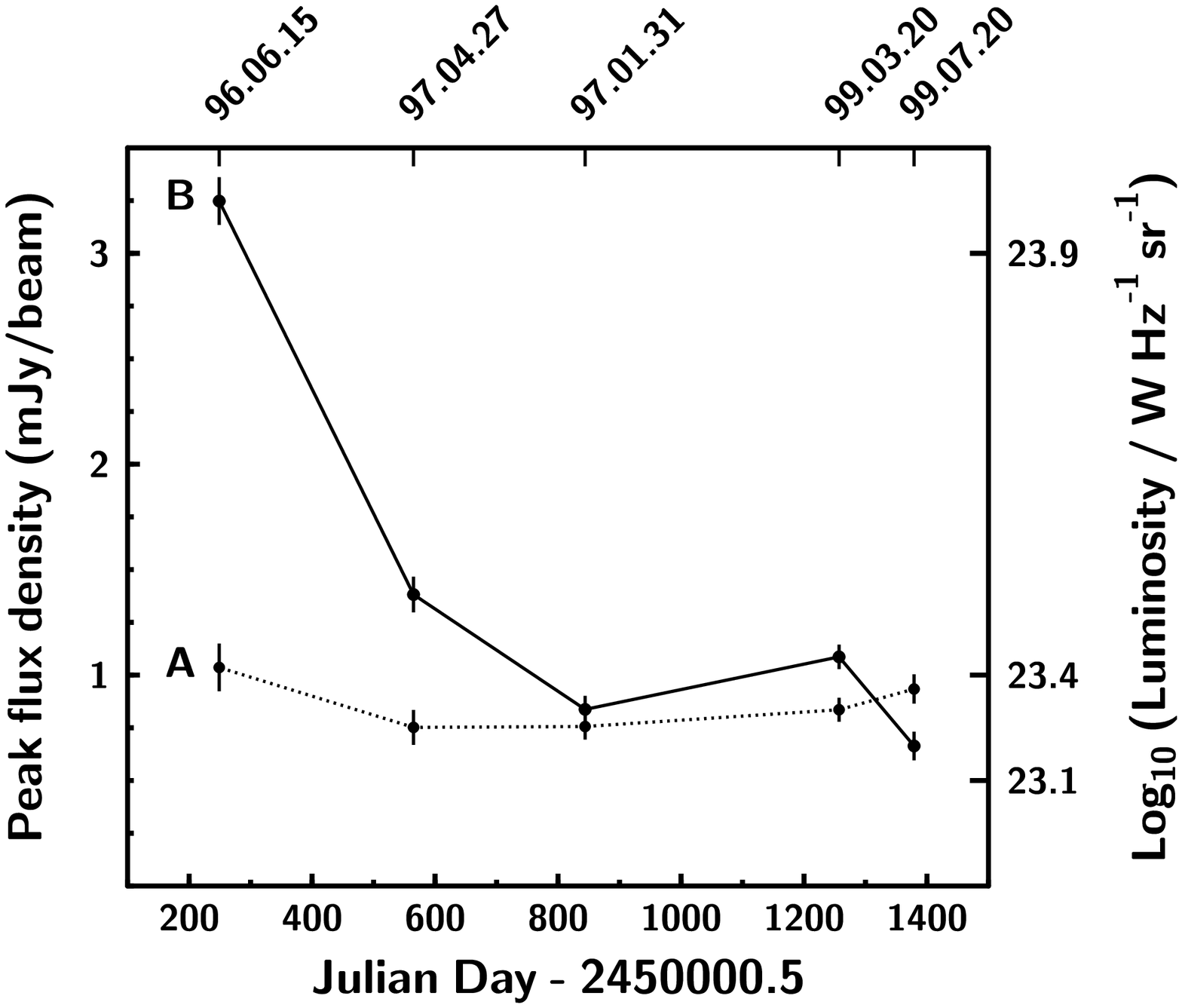,width=8cm,angle=0} 
\figcaption{\label{fig:fluxes}
Peak flux densities at 8.4\,GHz of components A \& B at each epoch
plotted against time.  The dotted line corresponds to the eastern
component (A) and the solid line corresponds to the western component
(B).  }

\section{Physical interpretation}
\label{s:interp}

It is feasible that the observed structural changes
in component B 
are related to at least one unresolved
sub-component. The much closer source 3C\,120 \citep{gomez00a} is a
good example of how jet knot components can appear and
fade.  G\'omez et al have interpreted this in terms of interaction
between the jet and the inhomogeneous ambient medium. Observed at
lower resolution, the jet in 3C~120 would present apparently
structurally changing partially resolved components. If we are indeed
observing such a phenomenon, then the offset between the two local
surface brightness maxima in component~B may give us some idea of the
scale of the jet diameter.  The perpendicular offset between the 
centroid of the two B
subcomponents and the line connecting the two main components, A and B
is about $ 0.7 \> \rm mas \approx 5.5 \> \rm pc$. Therefore, in the
following, we adopt a fiducial scale for the jet diameter, $D$ of $5
\> \rm pc$.  The causal limit on the diameter deduced from the
variability time scale, $\Delta t$, and corresponding to a Doppler
factor $\delta$ is:
$
D < c (1+z)^{-1} \, \Delta t \, \delta \approx 0.4
\left( \frac {\Delta t}{100 \> \rm days} \right) \,
\left( \frac {\delta}{10} \right) \> \rm pc.
\label{e:d_limit}
$ The variation from epochs 1 to 2 entails a factor $\sim 2$ decrease
in flux density over a timescale $\approx 400\> \rm days$. This
implies an upper limit on the diameter $\approx 1.6 \> \rm pc$ for a
Doppler factor of 10. This is intriguingly less than, but of the same
order as, the approximate scale size estimated from the VLBA images.
Of course, the discrepancy decreases with increasing Doppler factor.
[If the jet is pole-on so that $\delta \approx 2 \Gamma$ (where
$\Gamma$ is the bulk Lorentz factor), given the existence of
apparent proper motions $\sim 10 \, c$ in some quasars, then $\delta
\sim 20$ is not out of the question.]  Clearly, the reconciliation of
these two scales is in the direction of larger Doppler factors, not
smaller.  This is an indication that this source is highly beamed.

A second, more persuasive, indication for beaming comes from
estimating the intrinsic source power. The emitted power per steradian is:
$\int j^\prime_\nu dV^\prime = \, \delta^{-(3+\alpha)} 
\,
(1+z)^{-(1-\alpha)} \>  F_\nu \, D_L^2
\quad \rm W \> Hz^{-1} \> sr^{-1}$,
where the integral is that of the rest-frame emissivity,
$j^\prime_\nu$,  at the {\em observed} frequency $\nu$ over the proper 
volume ($V^\prime$) of each blob and $F_{\nu}$ is the flux density at 
$\nu$.

\psfig{figure=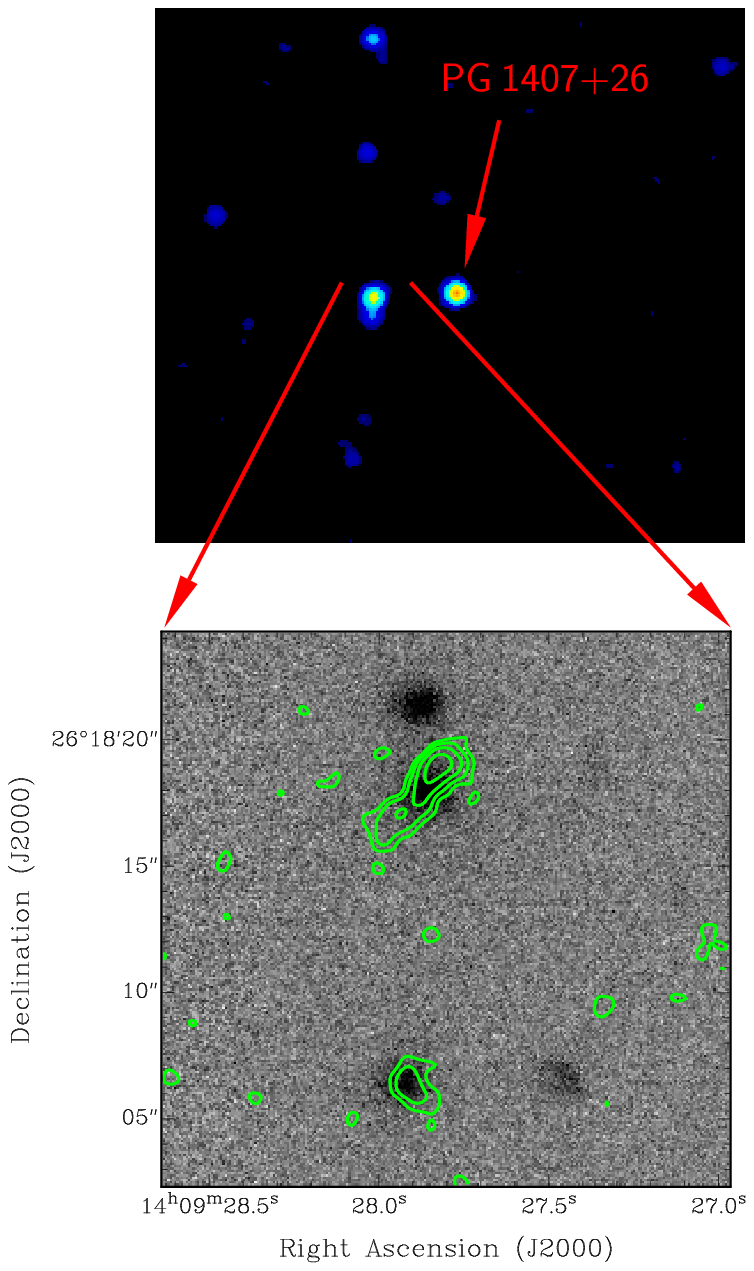,width=8cm,angle=0}
\figcaption{\label{fig:widefield} The top image is a seven
arcmin-square VLA image at 8.4\,GHz in D-array.  The lower image shows
a grey-scale of near-IR $K$-band on which are overlaid contours at
8.4\,GHz from the VLA A-array, the lowest of which is 0.04\,mJy/beam.
The two radio sources are clearly identified with resolved
18.5-magnitude galaxies, indicating they are independent radio
structures from the quasar.}

For unity Doppler factor, the total emitted power, at 8.4~GHz, of both
components at epoch~1 would be $1.1 \times 10^{24} \> \rm W \> Hz^{-1}
\> sr^{-1}$. This is comparable to the core powers of the most
powerful galaxies in the B2 sample \citep{deruiter90a}.  If
PG\,1407+263 were similar to the B2 sources then, using the
statistical relationship between core and extended powers derived by
\citet{deruiter90a}, its corresponding 1.4~GHz extended power would be
of order $10^{28} {\rm W \> Hz^{-1}}$ and the 8.4 GHz flux density
would be of order one Jansky.  However, since the flux density of any
extended emission from PG\,1407+263 is less than 0.1\,mJy, this is
another indication that the core is beamed.  The Doppler factor has to
be increased to $\sim 10$ in order to reduce the expected extended
flux density to the observed upper limit.
The additional parameters describing each knot are the spectral index
$\alpha$, the electron energy index, $a=2\alpha+1$, the volume,
$V^\prime$, of the knot in its rest frame, the Doppler factor,
$\delta$ and the ratio, $c_E$, of energy in relativistic electrons and
positrons to other particles. The rest-frame magnetic field that
minimizes the total energy density of a spherical blob in the rest
frame is: 
{\footnotesize
$B_{\rm min}^\prime = e^{-1} m_e \, \delta^{-1}(1+z)^{-\left(
   \frac{1-\alpha}{\alpha+3} \right)} \times$ \newline

$\left[ (a +1) \, C_2^{-1}(a) \, (1+c_E) \,  m_e^{-1} c
  \, f(a,\gamma_1, \gamma_2) \,
 F_\nu \nu^\alpha D_L^2{V^\prime}^{-1} 
\right]^{\frac{2}{a+5}}  \> {\rm T}.$}
Here $C_2(a)$ is a coefficient involving $\Gamma$-functions that
appears in the expression for the angle-averaged synchrotron
emissivity; $e$, $m_e$ and $c$ are the electronic charge, electronic
mass and speed of light respectively, and $f(a,\gamma_1, \gamma_2) =
(a-2)^{-1} \gamma_1^{2-a} (1-(\gamma_2/\gamma_1)^{2-a})$\footnote{See
http://www.mso.anu.edu.au/$\sim$geoff/HEA/HEA.html}.  We have
estimated the minimum-energy magnetic field and corresponding particle
energy densities, from the flux density of component B at epoch~1 and
for proper jet diameters of 0.375, 0.75 and 1.5\,mas as functions of
Doppler factor.  Figure\,\ref{f:energy} shows the associated jet
energy flux, $F_E \approx \left[ 4p \, \left( 1 + \frac
{\Gamma-1}{\Gamma} \frac {\rho c^2}{4p} \right) + \left( \frac
{{B^\prime}^2}{\mu_0}\right) \right] \, \Gamma^2 c \beta A_{\rm jet}$,
where $A_{\rm jet}$ is the cross-sectional area of the jet, $\rho$ is
the rest-mass density, and where the Poynting flux of a perpendicular
magnetic field has been included. We assume that the jet is pole-on
and purely relativistic, so that $\rho c^2/4p \ll 1$ and $c_E=0$.  The
almost constant value of the jet energy flux for Doppler factors $\ge
2$ is the result of the $\delta^{-2}$-dependence of the magnetic and
particle energy densities canceling the $\Gamma^2$-dependence of the
energy flux.  If we assume that 0.75\,mas is a reasonable upper limit
to the jet diameter, then the jet is no more powerful than about $7
\times 10^{36} \> \rm W$.  For jet diameters less than the fiducial
value the jet is less powerful, albeit still relativistic.
We note that this is not the first radio-weak object for which
relativistic velocities have been inferred. \citet{brunthaler00}
inferred superluminal motion from their observations of the Seyfert
galaxy IIIZw2, directly implying relativistic motion.

\psfig{figure=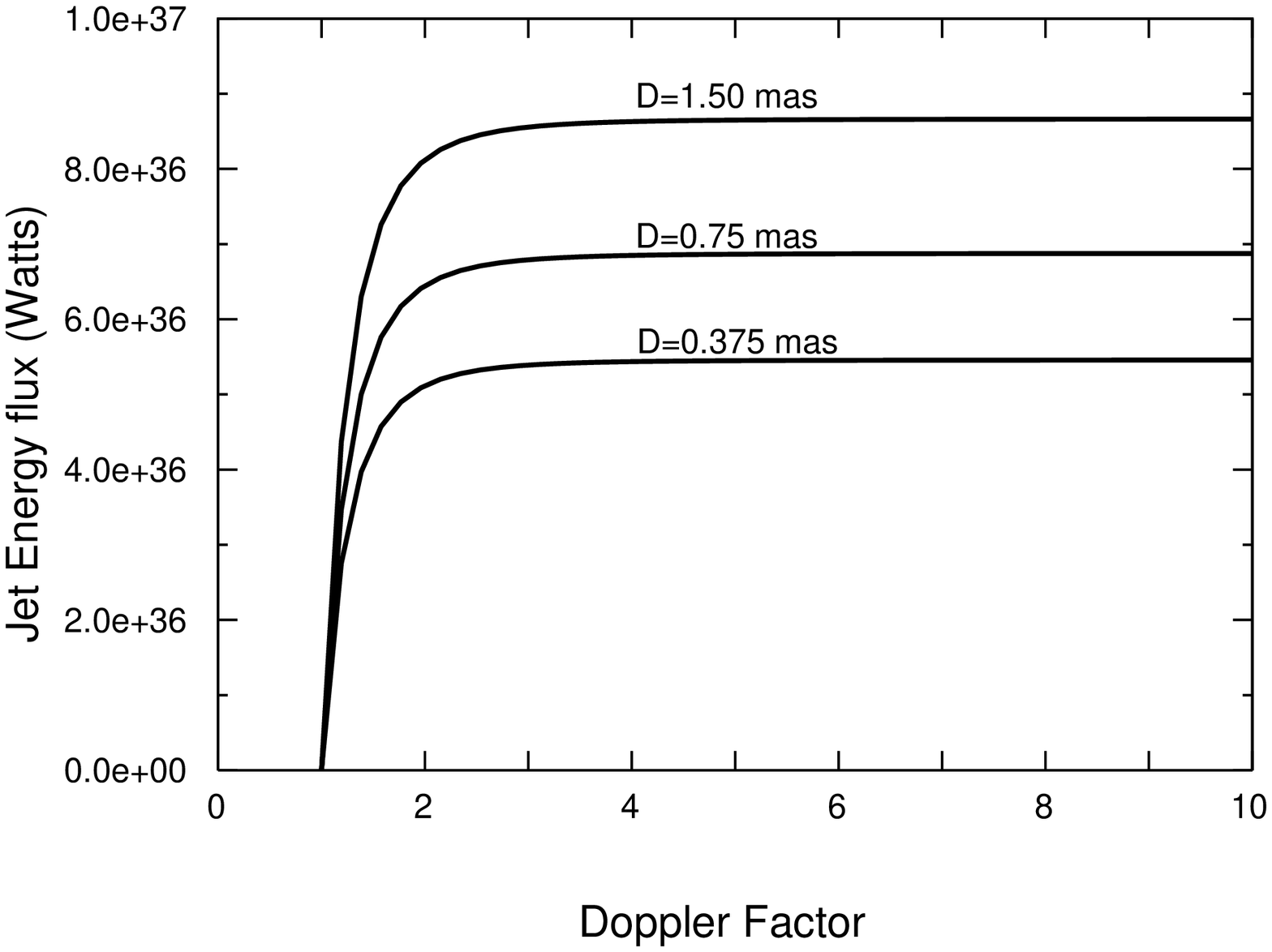,width=8cm,angle=0} 
\figcaption{\label{f:energy}
Jet energy flux as a function of Doppler factor for three assumed
values of the jet diameter: the fiducial jet diameter and a factor of
two above and below this. }

\section{Discussion}
The low extended emission relative to the core in PG\,1407+263 is a
good argument that this jet is relativistically beamed. If our
inference of partially resolved structure in the jet is correct then
the case for relativistic beaming is reinforced by the variability
time scale ($\sim 400 \> \rm days$ for epochs 1-2).  If the jet
diameter is indeed $\sim 5 \> \rm pc$ then, given the typical
expansion rates of jets, the components are of the order of 100\,pc
from the core, placing them in the narrow-line region. These two
arguments point to a Doppler factor of $\ga 10$.  As
Figure~\ref{f:energy} shows, the dependence of energy flux on Doppler
factor and jet radius indicates a kinetic power no greater than about
$10^{37} \> \rm W$. Using theoretical \citep{bicknell98a} and
observational \citep{willott99a} relations between jet and radio
power, such an energy flux would be typical of a radio galaxy jet no
more than an order of magnitude above the FRI/FRII break
\citep{Fan74}.  The interest that attends this galaxy is the existence
of an FRI radio power associated with a bright quasar so it is
unsurprising that this quasar is classified as radio quiet
\citep{Blu01}.  It appears the only reason that we detect any
radio flux is that it is one of the few PG quasars that is beamed
towards us.

As well as its variation between epochs 1 and 2, component B also
exhibits a significant rise in flux density $F_\nu$ between epochs 3
and 4 then a sharp drop ($\Delta t = 122$ days) between epochs 4 and 5
with $\Delta F_\nu/F_\nu \sim 1$. This variability indicates a
diameter $\le 0.5 (\delta /10) \> \rm pc $ for component B around this
epoch -- much less than the indicative jet diameter $\sim 5 \> \rm pc$
inferred above. There are two possibilities: (1) the jet may not be as
wide as 5\,pc or (2) the variation is due to a small section of the
jet being affected by its interaction with clouds in the narrow-line
region of the quasar.  This second case is similar to what has been
observed in 3C\,120 by \citet{gomez00a}.  Our observations therefore
suggest a relativistic jet, of only moderate power, propagating
through the clumpy narrow-line region of a bright host quasar.

Does our consideration of the radio structure of PG\,1407+263 shed any
light on its curious optical properties e.g., the low equivalent width
of the emission lines \citep{McD95}?  If the Doppler factor of the jet
is as high as 10, then we are viewing PG\,1407+263 within a few
degrees of pole-on and looking directly into its broad-line region.
Thus, the emission line spectrum may be diluted by a direct line of
sight to the accretion disk continuum. An argument against this is
that the PG quasars show no evidence for orientation-dependent effects
and this is generally ascribed to the initial selection via
ultra-violet excess.  However, supporting evidence for PG\,1407+263
being close to pole-on is the large velocity dispersion of its
emission line gas $\approx 10,000\> \rm km \> s^{-1}$ \citep{McD95},
indicating that we may be looking directly into the inner part of the
broad-line region. 
 
\acknowledgments

KMB thanks the Royal Society for a University Research Fellowship.
AJB gratefully acknowledges support from NSF grant AST-0116558
(Combined Array for Research in Millimeter-wave Astronomy).  The VLBA
and VLA are facilities of the NRAO operated by AUI, under co-operative
agreement with the NSF.  We would like to thank the UKIRT service
programme, and Dr Paul Hirst especially, for help in obtaining the
UFTI data.

\end{document}